\documentclass[aps,showpacs,twocolumn,superscriptaddress]{revtex4}
\usepackage{graphicx}
\usepackage{dcolumn}
\usepackage{amsmath}
\usepackage{color}

\begin{document}

\title{Quantum random number generator based on photonic emission in semiconductors}

\begin{abstract}
\rule{0ex}{3ex}
\noindent
{\bf
We report upon a novel realization of a fast nondeterministic random number generator whose randomness relies on intrinsic randomness of the quantum physical processes of photonic emission in semiconductors and subsequent detection by the photoelectric effect. Timing information of detected photons is used to generate binary random digits-bits. The bit extraction method based on the restartable clock method theoretically eliminates both bias and autocorrelation while reaching efficiency of almost 0.5 bits per random event. A prototype has been built and statistically tested.
\\

}
{\color{red} Cite as: M. Stip\v cevi\' c, B. Medved Rogina, "Quantum random number generator based on photonic emission in semiconductors", Rev. Sci. Instrum. {\bf 78}, 045104:1-7 (2007). DOI: 10.1063/1.2720728
}
\end{abstract}

\author{M. Stip\v cevi\' c}
\email{Mario.Stipcevic@irb.hr}
\affiliation{\footnotesize Rudjer Bo\v{s}kovi\'{c} Institute,
         Bijeni\v cka 54, P.O.B. 180, HR-10002 Zagreb, Croatia}

\author{B. Medved Rogina}
\affiliation{\footnotesize Rudjer Bo\v{s}kovi\'{c} Institute,
         Bijeni\v cka 54, P.O.B. 180, HR-10002 Zagreb, Croatia}

\maketitle

\section*{Introduction}

True random numbers, or more precisely nondeterministic random number generators, seem to be of an ever increasing importance. Random numbers are essential in cryptography (both classical and quantum), Monte Carlo numerical simulations and calculations, statistical research, randomized algorithms, lottery etc. 
\\

Historically, there are two approaches to random number generation: algorithmic (pseudorandom) and by hardware (nondeterministic). Pseudorandom number generators are well known in the art \cite{rngs}. A pseudorandom generator is nothing more than a mathematical formula which produces deterministic, periodic sequence of numbers which is completely determined by the initial state called {\em seed}. By definition such generators are not provably random. In contrast to that, hardware generators extract randomness form physical processes believed to behave in nondeterministic way which makes them better candidates for true random number generation. 
\\

In applications where provability is essential, randomness sources (if involved) must also be provably random. For example the famous BB84 quantum key distribution protocol described in \cite{bbbss91} would be completely insecure if only an eavesdropper could calculate (or predict) either Alice's random numbers or Bob's random numbers or both. From analysis of the secret key rate presented therein it is obvious that any guessability of random numbers by the eavesdropper would leak relevant information to him, thus diminishing the effective key rate. It is intriguing (and obvious) that in the case that the eavesdropper could calculate the numbers exactly, the cryptographic potential of the BB84 protocol would be zero. This example shows that the local random number generators assumed in BB84 are essential for its security and should not be taken for granted. 
\\

It has also been noticed that a random bit generator combined with an ordinary classical digital channel (this combination being referred to as "symmetric binary channel with noise") gives raise to cryptographic potential of such a channel \cite{wyner,ck,maurer}.
\\

\begin{figure}[h]
%\begin{figure}[p]
\centerline{\includegraphics[width=70 mm,angle=0]{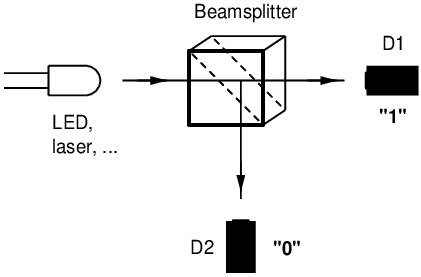}}
%\centerline{\includegraphics[width=150mm,angle=0]{splitter.eps}}
\caption{ Beam splitter is a frequently used component for random number generators. The two photon detectors D1 and D2 are used to detect two possible outcomes corresponding to one of the two possible paths that photon can take. Thus each photon entering the beam splitter generates one random binary digit - bit.}
\label{splitter}
\end{figure}

At present state of the art probably the best way to realize scientifically provable random number generator is to rely on intrinsic randomness of certain simple quantum systems. One such system, frequently exploited for random number generation, is based upon a photon passing through a beam splitter, as schematically shown in Fig. \ref{splitter}.
\\

A light source emits photons which pass through a beam splitter.
Whenever a photon is emitted it takes one of the two possible paths (with a small probability to get reflected or absorbed). If it ends up in the detector D1 then we generate binary value "1", whereas if it ends up in D2 we generate binary value "0". Because non-polarized light can be understood as an equal mixture of two orthogonal polarizations, there is a theoretical reason why zeros and ones would appear with the same probability in an idealized apparatus, that is why the bias (defined as $b=p(1)-1/2$ where $p(1)$ is probability of ones) would be zero. Other realizations of optical random quantum systems include: photons incident on a semi-transparent mirror \cite{rng-idquant}, polarized photons incident on a rotateable polarizing beamsplitter which offers a possibility to tune the bias \cite{jennewein1999} and an apparatus using polarization entangled photons to suppress false photon detections \cite{kinezi2004}. Practical realizations of quantum random number generators, of course, necessarily suffer from imperfections. Primary imperfection of such systems is that the probability of ones is not exactly equal to the probability of zeros. In the systems mentioned above this is firstly because photons which produce zeros and ones traverse different physical paths (which might have different transmission probabilities), and secondly because photons are detected with two (or more) detectors. Detection efficiency of single a photon detector can vary significantly form one physical device to another and is typically very sensitive to temperature drift, supply voltage fluctuations, component tolerances and aging. In order to achieve a low bias one needs to cope with two problems: first how to achieve that the detectors have exactly the same efficiencies (i.e. precisely such efficiencies as to compensate for different photon paths) and second how to keep the detectors efficiencies extremely constant. To minimize the bias, quantum random number generators described so far must be fine tuned for a low bias pr!
 ior to u
se. This is not convenient because this fine tuning is a very time consuming procedure due to the statistical nature of measurement of bias. Furthermore, due to stability problems one cannot expect that the bias once adjusted would stay constant for a long time.
\\

The source of problems with previous art is that it uses spatial information of photons randomly ariving at different places as a source of randomness. This necessarily requires two (or more) photon detectors. In our approach we use temporal information of photons emmited at random times by a light emitting diode or a laser. All photons traverse the same optical path and arrive always at the same single detector. Because now both zeros and ones are detected by the same detector, our random number generating method (described below) does not require any fine tuning of the bias. Furthermore the method is highly immune to stability problems because any slow change in the detector's performance influences by the same amount both zeros and ones and thus precisely cancels out. The price for improved randomness quality is paid in a somewhat more complex bit extraction method, but still easily manageable by logic electronics circuits. 
\\

The random number generator presented here consists of two principal parts: a physical random pulse generator (RPG) and a method of extracting random bits from such a generator. It will be shown that the method produces perfect random numbers (bits) if fed by perfectly random events. One practical realization will be presented and statitically tested.
 
\section*{Theory of operation: the restartable clock method}

For the purpose of this paper we define a random bit generator as a device which produces bits independently of each other and with equal probability of the two outcomes, i.e. $p(0)=p(1)=0.5$.
We assume existence of a physical setup which produces Poissonian random events, typically in form of digital pulses. By definition, Poisonian events are occuring independently of each other and in such a manner that time intervals between subsequent events follow the exponential probability distribution function (pdf).
\\

\begin{figure}[h]
%\begin{figure}[p]
\centerline{\includegraphics[width=70 mm,angle=0]{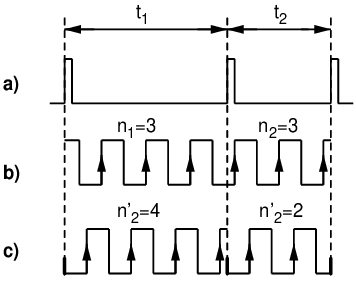}}
%\centerline{\includegraphics[width=150mm,angle=0]{qrbg-restart.eps}}
\caption{ Time interval measurement methods (a) with continuous (b) and restartable (c) clock. The time counter advances with every positive going edge of the clock pulse.}
\label{restart}
\end{figure}

Basic idea of the method for extracting random bits is to consider a pair of non overlapping random time intervals ($t_1$, $t_2$) which are defined with subsequent random events, as shown in Fig. \ref{restart}a, and generate either binary value "0" if $t_1 < t_2$, or "1" if $t_1 > t_2$. Next two intervals will be considered to generate the next random bit.
Since the events which determine time intervals are by definition independent of each other it is not possible that $t_1 < t_2$ would appear with any different probability than $t_1 > t_2$, consequently the probability to generate "0" is exactly equal to the probability to generate "1". In other words, the distribution of $t_{2i}-t_{2i-1}, i = 1,2,3,\ldots$ is symmetric.
Furthermore, the bits are mutually independent (i.e. uncorrelated) since independent pairs of events are used to generate different bits. 
\\

This "basic algorithm" readily (and obviously) yields perfect random bits if fed by perfectly random events. However, it needs to be modified in order to cope with common limitations of physical realizations, notably detectors and electronics, and still give good random numbers.
\\ 

First modification deals with the fact that measurement of time intervals usually results with an integer number which is produced, for example, by counting of a high frequency periodic signal (eg. quartz clock) or digizing a time to amplitude converter (TAC) output voltage. 
The time intervals $t_1, t_2$ are then represented by integer numbers $n_1, n_2$.
With this effect of "time discretization" it is still true that $p(n_1 < n_2) = p(n_1 > n_2)$ but now it can happen that $n_1 = n_2$. 
%% In fact, even without discretization, measurement of time can only be done with a finite precision which inevitably leads to a non zero probability that $t_1 = t_2$. 
In that case, in order to preserve equalness of probabilities of zeros and ones, the two intervals ($t_1$ and $t_2$) are simply discarded and no bit is generated.
\\

Next modification is crucial and deals with the fact that the time discretization leads to appearance of correlations among bits.
The serial autocorrelation coefficient with lag $k \geq 1$ of the sequence of bits $Y_1 ... Y_N$ is defined according to \cite{Knuth} as:
\begin{equation}
a_k = \frac{\sum_{i=1}^{N-k}(Y_i-\overline{Y})(Y_{i+k}-\overline{Y})}{\sum_{i=1}^{N}(Y_i-\overline{Y})^2}.
\label{eq:auto}
\end{equation}

Since different bits correspond to different random events we do not expect long-range correlations among bits. Therefore, we are mostly concerned with the lowest lag coefficient, $a_1$, which for simplicity we henceforth denote with $a$. 
\\

\begin{figure}[h]
%\begin{figure}[p]
\centerline{\includegraphics[width=88 mm,angle=0]{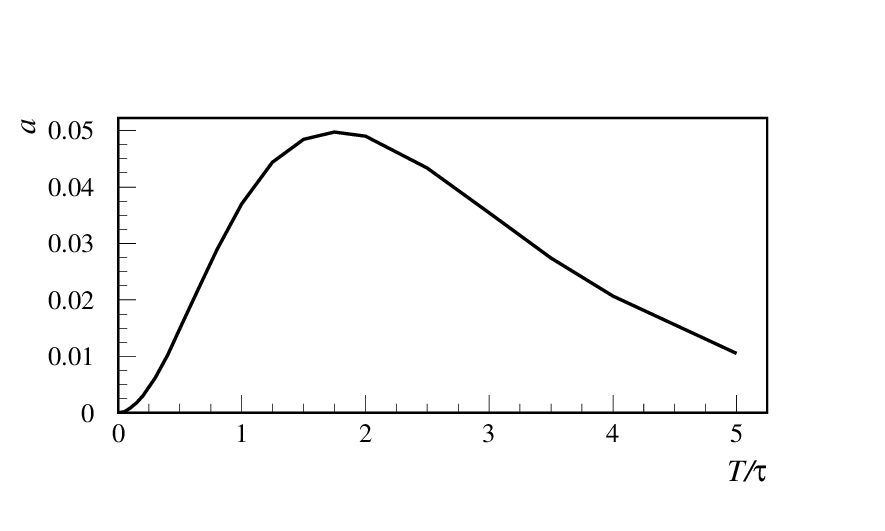}}
%\centerline{\includegraphics[width=150mm,angle=0]{anord.eps}}
\caption{ Serial autocorrelation coeficient $a$ for non-restartable clock method (dead time = 0). $T$ is the clock period, $\tau$ is the mean time interval between random pulses.}
\label{fig:anord}
\end{figure}

A technologically appealing way to measure (approximately) the time interval between subsequent random pulses is by counting periodic pulses from a continuous (i.e. non-restartable) quartz-controlled clock, as shown in Fig. \ref{restart}b. 
The time intervals $t_1, t_2$ are then represented by integer numbers $n_1, n_2$. Due to the symmetry between generating zeros and ones the method ensures zero bias. 
However since time intervals are only measured approximately it is not anymore obvious that bits obtained from such measurements will be random, as it was for the basic algorithm.
Indeed, simulations show that a sequence of bits will have a non-negative autocorrelation (Fig. \ref{fig:anord}) which vanishes in the limits of a very fast clock or a very slow clock.
This behavior can be understood as follows. 
%% Simulacije su izvedene programom extrac3 (extrac5 + grafika) -model nord

In the fast clock limit the time interval measurement becomes "exact" and this brings us back to the basic algorithm which provides unbiased and uncorrelated bits.
\\

In the slow clock limit the probability of having more than one count during a random period becomes negligible. Therefore bit "0" becomes mainly represented by $n_1=0, n_2=1$ whereas "1" becomes mainly represented by $n_1=1, n_2=0$. 
In other words, in the limit, one bit is generated with every clock count and 
if the clock count occurs during an even random interval then "0" is produced,
whereas if the clock count occurs during an odd random interval then "1" is produced.
During a clock period of length $T$, there are on average $T/\tau$ random intervals, where $\tau$ is the mean time interval between random pulses.
According to the central limit theorem,
the number of intervals becomes normally distributed as $T/\tau$ goes to infinity and since the normal distribution is symmetric even and odd periods (and thus 0's and 1's) appear with equal probability and at random. 
\\

\begin{figure}[h]
%\begin{figure}[p]
\centerline{\includegraphics[width=88 mm,angle=0]{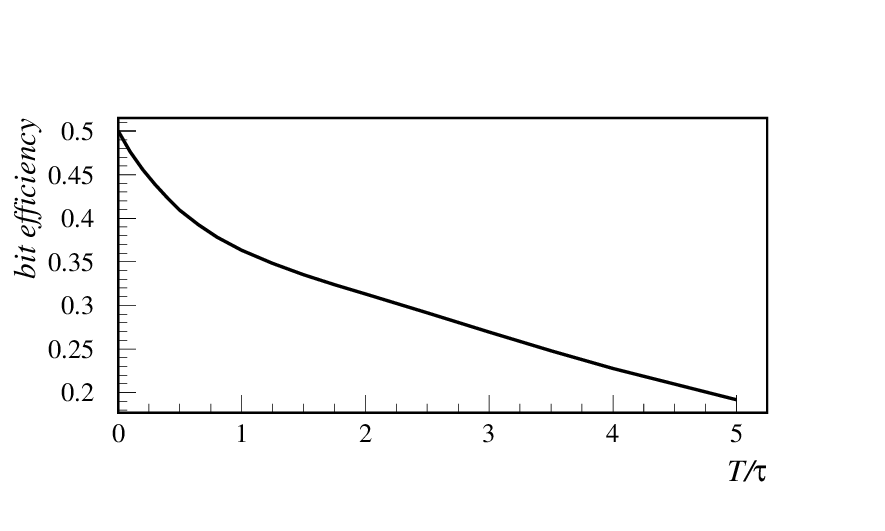}}
%\centerline{\includegraphics[width=150mm,angle=0]{enord.eps}}
\caption{ Bit efficiency, defined as the number of bits generated per random event, for non-restartable clock method as a function of $T/\tau$ where $T$ is the clock period and $\tau$ is the mean time interval between random pulses.}
\label{fig:enord}
\end{figure}

Apart from the correlation which characterizes the quality of randomness, another technologically important issue is the "bit efficiency"
%%, denoted $\eta_{bit}$, 
which we define as the number of
bits generated per random event. It reaches its maximum of 1/2 in the fast clock limit and continuously goes to zero at the slow clock limit (Fig. \ref{fig:enord}). Since good quality random events from the RPG are "technologically expensive", fast clock limit is preferred for practical realizations of random number generators.

Since the fast clock limit could never be reached in practice, it is interesting to understand what happens in its vicinity.
If we look at a pair of subsequently generated bits (for some $T/\tau > 0$), a process which involves measurements of 4 subsequent random time intervals, it turns out that this process slightly prefers pair values "11" and "00" (symmetrically) over the other two combinations, namely "01" and "10". 
This has a direct impact on autocorrelation since Eq. (\ref{eq:auto}) for $k=1$ is equivalent to:
\begin{equation}
a = p_{11}-p_{10}+p_{00}-p_{01} .
\label{altcorel}
\end{equation}

where $p_{11}$ is probability of "11" etc.
By simulation and symmetry arguments we have found that following relations hold for probabilities of pairs:
 $p_{11}$=$p_{00}$;
 $p_{10}$=$p_{01}$ and
 $p_{11} > p_{10}$
for finite $T/\tau$.
Acording to Eq. (\ref{altcorel}) this gives rise to a positive correlation.
Understanding of this autocorrelation effect from combinatorics and basic laws of probability is involved, but for our purpose it suffices to understand that the correlation appears as a consequence of the way in which random time intervals $t_i$ are measured. 
Figure \ref{restart}b illustrates this problem. If time interval $t_1$ ends up close to the beginning of a clock period it is more likely that $n_2$ will be smaller then in the case when $t_1$ ends up close to the end of a clock period. 
It means that measurement of the time interval $t_2$ depends on the length of $t_1$, that is $n_1$ and $n_2$ are somehow correlated.
\\

By help of extensive simulations we have derived an empirical asymptotic expansion for autocorrelation in the fast clock limit:
\begin{equation}
a(x) = \frac{4}{5} x^2
\label{fastclock}
\end{equation}

where $x = T/\tau$. This formula describes autocorrelation to better than 0.5% for $x \leq 0.2$. To see how this influences the bit production, let us assume we wish to have good quality random bits with an autocorrelation of the order of $10^{-4}$. According to Eq. (\ref{fastclock}), this is achieved if the ratio of the clock frequency and mean frequency of random events is about 90 ($x \approx 1/90$). For a very modest bit production speed of 2Mbit/sec this already requires clock frequency of 360MHz (bit efficiency $\approx 0.5$) which corresponds to fastest available logic circuits family.
The required minimum clock frequency scales linearly with the bit production speed and inverse of the square root of the desired autocorrelation which leaves no room for substantial improvement, either in the bit production rate or in correlation, with the present technology. Conversely, for a fixed clock frequency the autocorrelation scales {\em quadratically} with the desired bit production speed, which practically limits this method to bare 2Mbit/sec (250 kilobytes/sec). However autocorrelation can be completely removed by a different approach that we call "restartable clock".
\\

Namely, it is our finding that in order to avoid the correlations completely it suffices to synchronize clock pulses with beginning of each random interval, as shown in Fig. \ref{restart}c. 
In this case, result of measurement of each time interval is only a function of its length and obviously it is not dependent on any previous measurements, which completely eliminates correlations. Because all time intervals are measured in the same way, it is again true that $p(n_1 < n_2) = p(n_1 > n_2)$, meaning there is no bias either. The quality of radomness is now completely independent of ratio $T/\tau$. This "restartable clock method" theoretically yields perfectly random bits if fed by perfectly random events.
\\

The efficiency is again approximately described with the graph Fig. \ref{LeCroy}b. But now the departure from the maximum efficiency does not indicate any problem with the quality of randomness. It simply means that there are too few clocks within an average interval and that because of that we often have $n_1 = n_2$ in which case we must drop the whole pair of events. Good news is that with restartable clock method we can use clock of any speed: the only penalty for using a slow clock is efficiency lower than the maximum.
\\

It is important to understand here that the dead time caused by the photon counting system does not introduce correlations (nor bias) provided that the dead time following each detected photon is constant in length i.e. independent of any previously detected photons. This can be understood as follows. 
In the basic algorithm, the bit generating process makes use only of the difference of two time interval lengths. Since each random time interval contains exactly one dead time (namely the one following the first event) the two dead times cancel out completely. As a result, the distribution of time interval differences $t_{2i}-t_{2i-1}$ is independent of the dead time.
The same cancellation argument works also for restartable clock because the dead time is always measured in exactly the same way and produces always the same number of counts in the counter.
%% Provjereno programom cutexp.bas da je distribucija t_2i+1-t_2i neodvisna o mrtvom vremenu !
However the cancelation does not work for the non-restartable clock and indeed in that case (as we have checked by simulation) a non-zero dead time introduces additional correlation.

%% A last modification of the basic algorithm is that the third event in a triplet can be used as the first event in the next triplet in order to improve the bit efficiency. This does not introduce any correlations because we are only interested in time intervals, not the events themselves. 

\section*{Practical realization of the random pulse generator}

The random pulse generator is made of a photon source followed by a single photon detector. As a source of photons we have used a standard low-efficiency red light emitting diode (LED). LE diodes are direct bandgap devices which produce incoherent light by spontaneous emission \cite{Nakamura95} which is essentially a random process.
If operated at sufficiently low power, LED emits photons which are virtually independent of each other, that is the photon emission is then a Poissonian proces \cite{Nakamura95}. The important parameter here is the coherence time $\tau_{cohr}$, a time scale at which photons are becoming to be correlated. The coherence time can be estimated by help of the Heisenberg uncertainty relation, assuming Gaussian emission spectrum:

\begin{equation}
\tau_{cohr}=\frac{\lambda^2}{4\pi c \sigma}
\label{cohr}
\end{equation}

The type of LED we used has an exceptionally broad spectrum, approximately Gaussian 
centered at $\lambda =$ 688nm and width $\sigma \approx $ 35nm. 
If the mean frequency of photon emissions is kept well below the coherence frequency of $f_{cohr} = 2\pi/\tau_{cohr} \approx 2 \times 10^{15}$Hz, photons are practically mutually independent and consequently time intervals between subsequent photon emissions are exponentially distributed. Namely, the exponential distribution of time intervals between photon emissions and mutual independence of photons are two equivalent characterizations.
\\

Alternatively, instead of a LED diode one could as well use appropriately attenuated laser (diode). Namely, for a continuous, constant-power source of coherent light, a random distribution of photon emissions in time yields a Poissonian photon-counting distribution \cite{erdogan}. 
\\

Photons from the LED are directed toward photocathode of a photomultiplier tube.
Each time a photon hits the photocathode, there is a probability that an electron is emitted via the photoelectric effect. Creation of this so called "photoelectron" is a process governed by the binomial statistics. Therefore, the statistics of time intervals between emission of subsequent photoelectrons is governed by convolution of exponential and differential binomial distributions. However, the convolution of any distribution and exponential distribution gives again the exponential distribution. This is easy to understand  qualitatively: if we have a source of random events and choose randomly only some of these events, then the chosen events are again random. Even more favorably, the convolution of an approximately exponential distribution and any other distribution yields an improved exponential distribution, meaning that randomness of the photoelectric effect can only enhance overall randomness of produced photoelectrons.
%%As the matter of fact, the differential binomial distribution with $p$ = 0.001 is by itself almost an exponential distribution.
Therefore, creation of a photoelectron is virtually truly random event.
Creation of useable electric signal comprises: electron collection, 
electron multiplication, signal amplification, level detection and pulse shaping. The effect of this signal processing is that following a photon detection the detector becomes insensitive to further incoming photons during a certain period of time called the "dead time". For good quality detectors the dead time has a well defined, fixed length (which is important for our method). But even if this is not the case, that is if for a certain detector dead time varies from event to event, it is easy to modify such detector to achieve a constant dead time. This can be done, for example, by pasing its output through a non-retriggerable one-shot pulse shaping circuit which produces pulses of a constant length and which itself has the dead time slightly above the longest dead time observed in the detector. Such pulse shaping circuits, whose dead time can be set at a desired length, are well known in the art, see for example \cite{74HC221}. 
\\

In our experiments we have used Hamamatsu photomultiplier with bialkali photocathode R5611, high voltage source C3830, and the photon counting unit C3866 (amplifier + discriminator + pulse shaper). Dead time of this photon counting system is about 25 ns, very stable and independent of brightness of the LED. The low quantum efficiency of only about 0.05 percent of the bialkali photocathode at the wavelength of the LED presents no problem in our application. We have intentionally used low efficiency red LED because its widest wavelength spectrum in comparison to other LEDs indicates highest randomness of emited photons.
\\

\begin{figure}[h]
%\begin{figure}[p]
\centerline{\includegraphics[width=70 mm,angle=0]{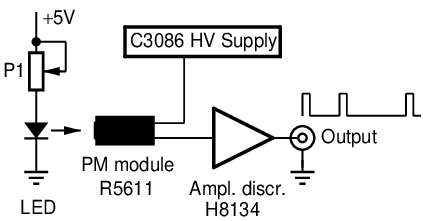}}
%\centerline{\includegraphics[width=150mm,angle=0]{gsp.eps}}
\caption{ Schematic diagram of the random pulse generator (RPG) composed of: light emitting diode (LED) and single photon detector setup. Mean frequency of output pulses can be set by potentiometer P1. }
\label{GSP}
\end{figure}

\begin{figure}[h]
%\begin{figure}[p]
\centerline{\includegraphics[width=80mm,angle=0]{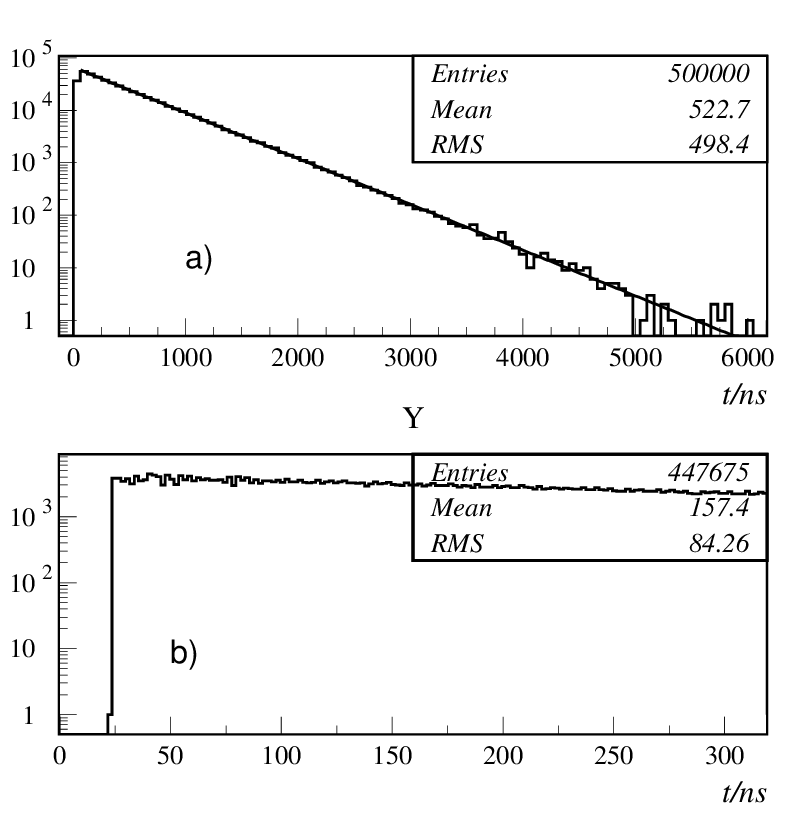}}
%\centerline{\includegraphics[width=150mm,angle=0]{raspod.eps}}
\caption{ (a) Measured probability density function (pdf) of time intervals between subsequent events from our setup; (b) zoom of the dead-time region which is characterized by a sharp transition from complete insensitivity of the detector following a photon detection to fully restored sensitivity some 25 ns later.}
\label{LeCroy}
\end{figure}

The probability density function (pdf) of measured time intervals between subsequent pulses from the RPG operated at 2 MHz as measured by the fast digital osciloscope LeCroy WaveRunner 6100A, is shown in Fig. \ref{LeCroy}a. An almost perfect fit of measured data to the exponential distribution extends over nearly five decades of the random variable. The effect of dead time is visible at the left end of the distribution as a sharp fall of the distribution at 25 ns. A zoom of this region is shown in Fig. \ref{LeCroy}b. Small wrinkles wisible at left end are caused by the binning effect due to the finite time resoution of the oscilloscope. 

Good agreement of the measured pdf with exponential distribution confirms that the photon emission of the LED is indeed random and that afterpulsing in the photon detector is negligible at the mean output frequency of 2 MHz. If, in some case, randomness of a single LED source would be suspected, one could use several independently driven LED sources to enhance the randomness of the light source. However, having in mind very small coherence time $\tau_{cohr}$ this is unlikely to be an issue in practice. On the other hand, the afterpulsing in photomultipliers could be a potential problem at high bit extraction rates. Afterpulsing in PMT's creates correlations in the distribution of random intervals which dies off tipically with a time constant of the order of $\tau_{corr} \approx 1000$ ns \cite{pmtafterpuls}. At large photon incidence frequencies this effect becomes negligible in comparison with abundant signal caused by real photons. Nevertheles, in principle, afterpulsing could cause serial correlations among random bits when bit extraction rate exceeds $1/\tau_{corr}$ no matter what the extraction method. This effect should be studied in more detail preferably by simulations which greatly over emphasize the level of afterpulsing. In any case, bits produced by our prototype at 1 Mbit/sec do not show any measurable correlations, as will be presented later.
\\

The non-zero dead time, as will be shown, does not create bias nor correlations among random bits in our method.

\section*{The random number generator prototype}

\begin{figure}[h]
%\begin{figure}[p]
\centerline{\includegraphics[width=80 mm,angle=0]{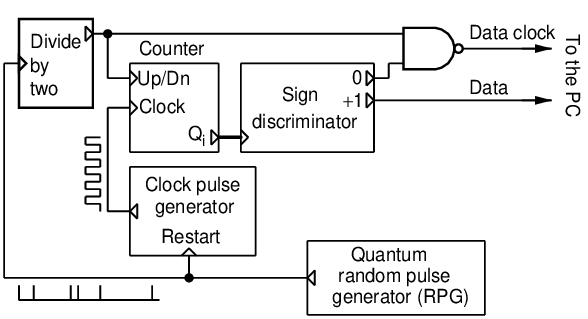}}
%\centerline{\includegraphics[width=150mm,angle=0]{qrbg-block.eps}}
\caption{ Diagram of the quantum random number generator consistsing of random pulse generator and  curcitry for implementation of the restartable clock method.}
\label{QRBG}
\end{figure}

Using the above RPG and logic circuitry embedding our bit extraction method 
we have built and tested a laboratory prototype of a quantum random number generator which operates upon the above described method. Big advantage of this method is that it requires only one photon detector and is insensitive to slow variations in the detector's gain and quantum efficiency, in contrast to some other methods which crucially depend on extreme stability of these parameters over a full lifetime of the detector.
Simplified schematic diagram of the setup is shown in Fig. \ref{QRBG}. It contains a quantum random pulse generator described in the previous section (Fig. \ref{GSP}).
%% Nacrtati shemu s up/down counterom kao u patentu

For the construction of the prototype we have used a controlable bi-directional (up/down) counter which operates in the following way.
At the beginning of any first time interval, in a pair of subsequent random time intervals as explained earlier, the counter state is set to 0.
During each first random time interval its control input is held at high logic state so it counts forward, and during the second random interval the control input is set low, so it continues counting backward until the end of the second random time interval. The required difference $n_1-n_2$ is thus effectively obtained without a need to actually perform any subtraction. If the counter state represents negative number we generate "0", if positive we generate "1" and if zero we generate nothing.
\\

Implementing the extraction algorithm in the discrete 74HC logic family and operating the random event generator 
%%(similar to the one shown in Fig. \ref{GSP}) 
at a mean frequency $f_0 = 2$MHz, we were able to extract random bits at almost 1 million bits per second. 
With a 48MHz restartable clock we have measured bit efficiency of $0.487 \pm 0.02$, which is quite close to the theoretical limit of 0.5, and in a good agreement with theoretical prediction of 0.490. Namely it can be shown that the bit efficiency of our method $\eta_{bit}$, defined as the number of bits produced on average per 1 random event asymptotically behaves like:
\begin{equation}
\eta_{bit} = \frac{1}{2} - \frac{1}{4}x + \frac{1}{8}x^2 \pm O(x^3)
\end{equation}

where $x = T/\tau$. The error of this asymptotic expansion is less than 0.25 percent for $x \leq 0.5$.
Since in our case $x$=2/48 this above formula gives $\eta_{bit} \approx 0.490$.
\\

Although theoretically eliminated, the main problem with the practical realization of this prototype was a non-vanishing bias. This seems to be a general problem with physical random number generators. In our case the bias was probably mainly due to the fact that switchings of the bi-directional counter to "up" direction and to "down" direction are not equally fast. 
Even the slightest systematic difference in the length of "up" and "down" counting time intervals has a measurable effect on bias. 
It can be shown that in the case of zero dead time the bias is given by:
\begin{equation}
%% b = \frac{1}{2} \left( \frac{T}{\tau} \right) (t_{up->down} - t_{down->up})
b = \frac{1}{2} \frac{T}{\tau} \frac{\Delta t}{\tau}
\label{finetune}
\end{equation}

where $\Delta t$ is the difference between transition times for "up" and "down" counting directions.
However this biasing effect has no bearing on correlations which were indeed consistent with zero.
 
The bias of this setup was measured to be very small, of the order of $10^{-4}$. This is readily by far the smallest bias reported for the raw bits from a quantum random bit generator \cite{quantis,jennewein1999}. However, bias quickly raised with random events freqency arriving to $10^{-3}$ at $f_0 \approx 6$MHz, the scaling with $1/\tau^2$ being quite in accordance with Eq. (\ref{finetune}).
%% In our case bias appears because the time periods for counting up and down are not exactly of the same length. 
\\

In order to lower the bias even further or make possible operation at higher random events frequencies one could add a small tuneable delay to the clock input of the up/down counter and adjust it for a minimum bias.

Alternative approaches include use of faster electronics which has smaller internal circuit delays and software debiasing procedures.

\section*{Statistical quality of random numbers}

Quality of randomness of long sequences of bits produced by our prototype was evaluated by three batteries of statistical tests: ENT \cite{ent}, DIEHARD \cite{DIEHARD} and STS version 1.5 for Linux \cite{sts-1.50}. 

ENT is a collection of "standard" basic statistical tests, including autocorrelation test.
We have modified this program in order to be able to check correlation coefficients with lags higher than 1. Notably, we have checked serial correlation coefficients (Eq. \ref{eq:auto}) with lags from 1 up to 32 to be consistent with zero within statistical error bars which were of the order of $10^{-4}$.
Below is a result for a typical sequence of 1E9 bits. For sake of brevity only the first 8  serial correlation coefficients are shown in  Table \ref{entresults}.
\\

\begin{table}[h]
\begin{tabular}{l}
Entropy = 1.000000 bits per bit. \\
Chi square distribution for 1E9 samples is 2.89, \\
and randomly would exceed this value \\
25.00 percent of the times. \\
Arithmetic mean value of bits is   0.500086 +/- 0.000050 \\
Monte Carlo value for Pi is 3.140953 (error  0.02 percent). \\
Serial correlation coef. $a_01$ = -0.000027 +/- 0.000100 \\
Serial correlation coef. $a_02$ =  0.000092 +/- 0.000100 \\
Serial correlation coef. $a_03$ =  0.000101 +/- 0.000100 \\
Serial correlation coef. $a_04$ =  0.000131 +/- 0.000100 \\
Serial correlation coef. $a_05$ = -0.000116 +/- 0.000100 \\
Serial correlation coef. $a_06$ =  0.000121 +/- 0.000100 \\
Serial correlation coef. $a_07$ =  0.000082 +/- 0.000100 \\
Serial correlation coef. $a_08$ = -0.000025 +/- 0.000100 \\
\end{tabular}
\caption{Results of ENT tests suite for a typical sequence of 1E9 bits.}
\label{entresults}
\end{table}

We can see a slight tendency toward positive bias, also indicated by larger Chi square. Running this test over several sequences we have concluded that the bias of our generator is of the order of 1E-4. However, correlations were always consistent with zero.
\\

DIEHARD is a renown battery of tests which is most sensitive to various problems likely to appear in pseudorandom generators. It is widely considered as a most stringent series of statistical tests. It consists of 15 tests whose outcome is a p-value. According to the acompanying documentation, a test is considered failed if the corresponding p-value is less than 0.00001 or greater than 0.99999. Some tests have variants so all in all there are 20 p-values. Results of testing of the sequence of 1E9 bits shown in the Table \ref{dieresults} indicate that all the tests were passed.

\begin{table}[h]
\begin{tabular}{ll}
\hline
 Birthday Spacings            & 0.465507   \\ 
 Overlapping Permutations     & 0.12892    \\  
 Ranks of 31x31 Matrices      & 0.57825    \\ 
 Ranks of 32x32 Matrices      & 0.795461   \\     
 Ranks of 6x8 Matrices        & 0.862294   \\   
 Monkey Tests on 20-bit Words & 0.72331    \\ 
 Monkey Test OPSO             & 0.9383     \\ 
 Monkey Test OQSO             & 0.6845     \\ 
 Monkey Test DNA              & 0.1439     \\   
 Count 1's in Stream of Bytes & 0.146824   \\ 
 Count 1's in Specific Bytes  & 0.290347   \\ 
 Parking Lot Test             & 0.554244   \\ 
 Minimum Distance Test        & 0.386845   \\        
 Random Spheres Test          & 0.375526   \\ 
 The Sqeeze Test              & 0.667035   \\ 
 Overlapping Sums Test        & 0.859314   \\ 
 Runs Test (up)               & 0.173183   \\  
 Runs Test (down)             & 0.137973   \\ 
 The Craps Test no. of wins   & 0.176271   \\ 
 The Craps Test throws/game   & 0.523654   \\ 
\hline
\end{tabular}
\caption{Results of DIEHARD tests suite for a typical sequence of 1E9 bits.}
\label{dieresults}
\end{table}

NIST has recently put forward "A statistical test suite for random and pseudorandom number generators for cryptographic applications" (STS) which aims to become first industry standard for testing of random number generators. STS tests consist of total of 38 individual statistical tests, such as entropy, chisquare, long runs, DNA tests, birthday spacing tests, Lempel-Ziv complexity, Maurer's test etc. 
The written output from STS is lengthy and therefore not suitable for presenting here. It suffices to say that we have repeatedly tested sequences as long as $1 \times 10^9$ bits, which passed all of the tests.

\section*{Conclusion}

We have presented the concept and the development of a new type of a fast nondeterministic random number generator whose randomness relies on intrinsic randomness of the quantum physical process of photonic emission in semiconductors and subsequent detection by the photoelectric effect. Timing information of detected photons, or more precisely the time interval between detected random events, is used to generate binary random digits-bits.
\\

Presented prototype consist of a light source, one single-photon detector and fast electronics for timing analysis of the detected photons providing random output numbers (bits) at 1Mbit/sec. By using only one photodetector there is no need to perform any fine tuning of the generator, moreover method is immune to detector instability problems, which are improved characteristics compared to most of the systems based on the spatial photon detection used today. Another originality is that for the purpose of eliminating correlations        time interval measurement principle has been modified to a restartable clock method. The collection of statistical tests applied to random numbers produced with our quantum random number generator present results which demonstrate the high quality of randomness resulting in bias less than $10^{-4}$ and autocorrelation consistent with zero, with efficiency of $0.487 \pm 0.02$ bits per detected random event. 
\\

With appearance of fast quantum random generators in the market \ref{} and obvious declining of speed of computer processors, the speed gap between software and hardware generators has become much smaller especially if one takes into account that hardware generators do not put such a load on the processor as software generators do. We predict that quantum random number generators will find their applications in basic scientific research, sensitive data protection and encryption, as well as in simulations in science and technology. 
\\

We are grateful to prof. B. Basrak form the Mathematical faculty of the University of Zagreb for enlightenig discussions.
Parts of this work are subject of the patent application WO2005106645. Building of the prototype was supported by World Bank Technical Assistance Program 2004 for Croatia (TAL-2).

\pagebreak
\section{Figure Captions}

Fig. 1.  Beam splitter is a frequently used component for random number generators. The two photon detectors D1 and D2 are used to detect two possible outcomes corresponding to one of the two possible paths that photon can take. Thus each photon entering the beam splitter generates one random binary digit - bit.
\\

Fig. 2. Time interval measurement methods (a) with continuous (b) and restartable (c) clock. The time counter advances with every positive going edge of the clock pulse.
\\

Fig. 3. Serial autocorrelation coeficient $a$ for non-restartable clock method (dead time = 0). $T$ is the clock period, $\tau$ is the mean time interval between random pulses.
\\

Fig. 4. Bit efficiency, defined as the number of bits generated per random event, for non-restartable clock method as a function of $T/\tau$ where $T$ is the clock period and $\tau$ is the mean time interval between random pulses.
\\

Fig. 5. Schematic diagram of the random pulse generator (RPG) composed of: light emitting diode (LED) and single photon detector setup. Mean frequency of output pulses can be set by potentiometer P1. 
\\

Fig. 6. (a) Measured probability density function (pdf) of time intervals between subsequent events from our setup; (b) zoom of the dead-time region which is characterized by a sharp transition from complete insensitivity of the detector following a photon detection to fully restored sensitivity some 25 ns later.
\\

Fig. 7. Diagram of the quantum random number generator consistsing of random pulse generator and  curcitry for implementation of the restartable clock method.

\end{document}